\documentclass[aps,prl,preprint,superscriptaddress,floatfix]{revtex4-1}
\usepackage[utf8]{inputenc}
\usepackage{graphicx}
\usepackage[dvipsnames]{xcolor}
\usepackage[breaklinks=true,colorlinks=true,allcolors=Blue]{hyperref}
\usepackage{amsmath,amssymb}
\usepackage[T1]{fontenc}
\usepackage{newtxtext}\usepackage[varg]{newtxmath}
\usepackage[capitalize]{cleveref}

\begin{document}
\title{Tracing the Pathway from Drift-Wave Turbulence with Broken Symmetry to the Production of Sheared Axial Mean Flow}
\date{\today}
\author{R.~Hong}
\affiliation{Center for Energy Research, University of California San Diego, La Jolla, CA 92093, USA}
\author{J.C.~Li}
\affiliation{Center for Astrophysics and Space Sciences, University of California San Diego, La Jolla , CA 92093, USA}
\author{S.~Chakraborty~Thakur}
\author{R.~Hajjar}
\affiliation{Center for Energy Research, University of California San Diego, La Jolla, CA 92093, USA}
\author{P.H.~Diamond}
\affiliation{Center for Astrophysics and Space Sciences, University of California San Diego, La Jolla , CA 92093, USA}
\affiliation{Center for Fusion Science, Southwestern Institute of Physics, Chengdu, Sichuan 610041, China}
\author{G.R.~Tynan}
\affiliation{Center for Energy Research, University of California San Diego, La Jolla, CA 92093, USA}
\affiliation{Center for Fusion Science, Southwestern Institute of Physics, Chengdu, Sichuan 610041, China}

\begin{abstract}
This study traces the emergence of sheared axial flow from collisional drift wave turbulence with broken symmetry in a linear plasma device---CSDX.
As the density profile steepens, the axial Reynolds stress develops and drives a radially sheared axial flow that is parallel to the magnetic field.
Results show that the non-diffusive piece of the Reynolds stress is driven by the density gradient and results from spectral asymmetry of the turbulence and thus is dynamical in origin.
Taken together, these findings constitute the first simultaneous demonstration of the causal link between the density gradient, turbulence and stress with broken spectral symmetry, and the mean axial flow.
\end{abstract}

\maketitle

Symmetry breaking plays a vital role in pattern formations, in particular the generation of macroscopic flow by turbulence \cite{Shtern1999ARoFM537,Diamond1994PRL2565,Wu2017S}.
In stars and planetary atmospheres, convection driven turbulence drives macroscopic flows of interest via Reynolds stresses.
Broken symmetry defines the cross-phase (coherence) of the turbulent stresses, and thus is fundamental to the flow structures.
Examples of flow generation mechanisms exploiting broken symmetry include, but are not limited to, the anisotropic kinetic $ \alpha $-effect \cite{Frisch1987PDNP382} and $\Lambda$-effect \cite{Ruediger1989}.
The need for such broken symmetry also appears in the magnetic dynamo, where
turbulence with broken reflection symmetry (i.e., helicity) is required to produce a mean field dynamo \cite{Moffatt1978}.

The modelization of intrinsic macroscopic flow in plasmas also involves symmetry breaking in the turbulence.
Intrinsic flow is of great importance in magnetic confinement fusion due to its promise of stabilizing MHD instabilities \cite{Garofalo2002PRL235001} and reducing turbulent transport \cite{Yoshida2008PRL105002,Rice2016PPaCF83001}, particularly in burning plasma devices like ITER, where momentum input is limited.
Measurements from Alcator C-Mod \cite{Rice2011PRL215001} indicate that intrinsic toroidal flow in H-mode is driven by edge $ \nabla T $.
One possible mechanism \cite{Rice2011PRL215001,Kosuga2010PoP102313} is that free energy stored in radial gradients is converted into shear flows via underlying turbulent stress, analogous to a heat engine process \cite{Ozawa2003RoG}.
In this mechanism, broken symmetry in spectra of drift wave turbulence, $\langle k_\theta k_z \rangle \neq 0$, induces a residual, non-diffusive component ($ \Pi_{rz}^{\text{Res}} $) in the Reynolds stress ($ \left\langle \tilde{v}_{r}\tilde{v}_{z}\right\rangle = -\chi_{z} \partial_r V_{z} + V_{\text{p}} V_z + \Pi_{rz}^{\textrm{Res}} $) \cite{Kosuga2010PoP102313,Diamond2013NF104019}.
The divergence of this residual stress then defines an intrinsic torque that drives the macroscopic shear flow.

While many observations manifest the correlation between macroscopic intrinsic flow and edge profile gradients \cite{Ida2014NF45001,Rice2016PPaCF83001}, investigations of the microscopic mechanism have been limited.
Probe measurements from the edge of TJ-II \cite{Goncalves2006PRL145001} and TEXTOR \cite{Xu2013NF72001} suggest that the non-diffusive residual stress, $ \Pi_{rz}^\textrm{Res} $, scales linearly with edge gradients.
Parallel flow driven by turbulent Reynolds stress has been observed in a linear device, PANTA \cite{Inagaki2016SR22189}.
A recent gyrokinetic simulation has predicted a dipole structure for $ \Pi_{rz}^{\text{Res}} $, implying an intrinsic torque that is consistent with the measured rotation profile in DIII-D \cite{Wang2017PoP92501}.
Despite of all these advances, our understanding of the underlying physics is still far from complete.
Until now, there is no direct evidence linking the turbulence symmetry breaking to the development of residual stress.
Moreover, it is still not clear if the residual stress can efficiently convert the thermodynamic free energy into  the kinetic energy of the intrinsic flow.

In this study, we address the fundamental physics of how intrinsic flow develops in a confined plasma without magnetic shear.
The results presented here constitute the first experimental evidence that \emph{macroscopic radially sheared parallel flows develop from a finite residual stress which emerges from a dynamical symmetry breaking in the spectral correlator} $ \left< k_{z} k_{\theta} \right> $ of collisional drift wave turbulence.
In this case, the symmetry breaking \cite{Li2016PoP52311} is \textit{not} due to magnetic geometry, but rather due to a phenomenon similar to modulational instability, in which a small seed axial shear is amplified by the turbulence.
Note that modulational instability is also responsible for the production of zonal flows.
However, zonal flow generation does not require broken symmetry in the turbulence spectra \cite{Diamond2005PPaCF35}.
Thus, the mechanism for the generation of axial flow is more delicate than that for azimuthal flow.

The experiments have been conducted on the Controlled Shear Decorrelation eXperiment (CSDX), a cylindrical plasma device with an overall length of 2.8 m and a diameter of 0.2 m \cite{Burin2005PoP52320,Thakur2014PSSaT44006}. 
The working gas is argon at a gas fill pressure of 1.8 mTorr.
This relatively lower neutral pressure is used to avoid the volumetric recombination and detachment phenomena.
The argon plasma is produced by a 13.56 MHz 1800 W RF helicon source via an $m=1$ antenna, and is terminated by insulating (glass) end-plates at both ends.
The uniform magnetic field is in the axial direction (denoted as the $\hat{z}$ direction) and is scanned from 500 G to 1000 G in this study.
The peak electron temperature is about 4 eV, and the peak ion temperature is about 0.5 eV.
More details on this device can be found in previous publications \cite{Burin2005PoP52320,Thakur2014PSSaT44006}.

Plasma velocities is measured using a combined Langmuir and Mach probe array.
Two Mach probe tips, which are aligned along the axial direction, are used to measure the axial flow. 
The axial velocity is given by $ v_{z}=Mc_{s}=0.45c_{s}\ln\left(\frac{J_{\textrm{u}}}{J_{\textrm{d}}}\right) $,
where $c_{s}=\sqrt{T_{e}/m_{i}}$ is the sound speed and $J_{\textrm{u,d}}$ are the ion saturation fluxes collected by Mach probe tips at the up- and down-stream side.
The probe geometry is small enough to avoid shadowing effects that lead to spuriously large inference of parallel velocities \cite{Gosselin2016PoP73519}.
The mean parallel velocities found with this Mach probe are consistent with published measurements made using laser induced fluorescence diagnostics \cite{Thakur2016PoP82112}.
The fluctuating $\mathbf{E\times B}$ velocities are estimated from the floating potential gradients between two adjacent tips ($\nabla\tilde{\phi}_{\textrm{f}}$), i.e.,  $\tilde{v}_{r}=-\nabla_{\theta}\tilde{\phi}_{\textrm{f}}/B$ and $\tilde{v}_{\theta}=\nabla_{r}\tilde{\phi}_{\textrm{f}}/B$. 
The distance between two adjacent floating potential tips is about 3 mm.
The sampling rate of the probe data is $f_{s}=500$ kHz, which is well above the frequency of the observed dominant fluctuations ($f<30$ kHz) \cite{Thakur2014PSSaT44006}.
Similar probe configurations have also been used in other studies on the structure of parallel flows \cite{Inagaki2016SR22189}.

\begin{figure}
    \includegraphics[width=3.3in]{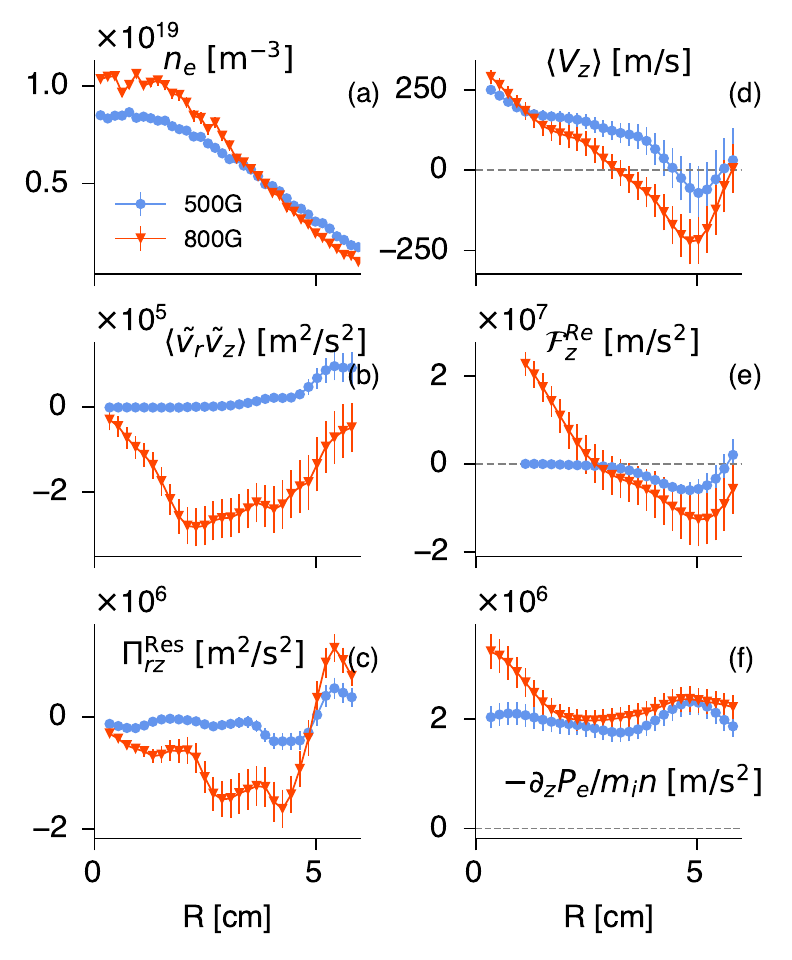}
    \caption{\label{fig:profiles} Equilibrium profiles of (a) plasma density $ n_e $, (b) parallel Reynolds stress $ \langle \tilde{v}_r \tilde{v}_z \rangle $, (c) Residual stress $ \Pi_{rz}^\textrm{Res} = \langle\tilde{v}_{z}\tilde{v}_{r}\rangle + \langle \tilde{v}_r^{2} \rangle \tau_{c} \partial_r V_z $, (d) axial velocity, (e) Reynolds force $ \mathcal{F}_{z}^{Re} = - \partial_r \langle\tilde{v}_{z}\tilde{v}_{r}\rangle $, (f) force on ions due to axial electron pressure gradient $ -\frac{\partial_z P_e}{m_i n} $.}
\end{figure}

In this study, we obtained different equilibrium profiles and fluctuation intensities by varying the $ B $ field.
As shown in \cref{fig:profiles}(a), when the $B$ field is raised, the plasma density profile \emph{steepens}.
During the $ B $ scan, the variation in electron temperature profiles is negligible.
The axial Reynolds stress, $\langle\tilde{v}_{z}\tilde{v}_{r}\rangle$, is estimated using velocity fluctuations in the frequency range of $2<f<30$ kHz.
Previous studies have identified these as resistive drift wave fluctuations \cite{Burin2005PoP52320}.
$\langle\tilde{v}_{z}\tilde{v}_{r}\rangle$ is negligible for $r<3$ cm at lower $B$ field, but becomes substantially negative at higher $B$ field (\cref{fig:profiles}(b)).
The Reynolds force, $ \mathcal{F}_{z}^{Re} = - \partial_r \langle\tilde{v}_{z}\tilde{v}_{r}\rangle $ (\cref{fig:profiles}(e)), increases significantly in the core and becomes more negative at the edge ($ 3<r<6 $ cm), and the radial shear of axial flow gets stronger as the $B$ field increases (\cref{fig:profiles}(b)).
The residual stress (described in more detail later) is computed from measured quantities using $ \Pi_{rz}^\textrm{Res} = \langle\tilde{v}_{z}\tilde{v}_{r}\rangle + \langle \tilde{v}_r^{2} \rangle \tau_{c} \partial_r V_z $ \cite{Yan2010PRL65002}, where $ \tau_{c} $ is the eddy correlation time.
Note that the momentum pinch ($ V_{\textrm{p}}V_{z} $) vanishes due to the lack of toroidal effects in CSDX.
The magnitude of the resulting $ \Pi_{rz}^\textrm{Res} $ also increases as $B$ field is raised (\cref{fig:profiles}(c)).
At $ B=800\, \textrm{G}$, the axial Reynolds force is much larger than the force on the ions arising from the parallel electric field.
Here, Boltzmann equilibrium is assumed and the weak electric field is inferred from the measured electron pressure drop along the axial direction, $ -\partial_z P_e /m_i n $ (\cref{fig:profiles}(e)), which is measured by two Langmuir probes at up- and down-stream locations ($\Delta z = 1.5$~m).

\begin{figure}
    \includegraphics[width=3.3in]{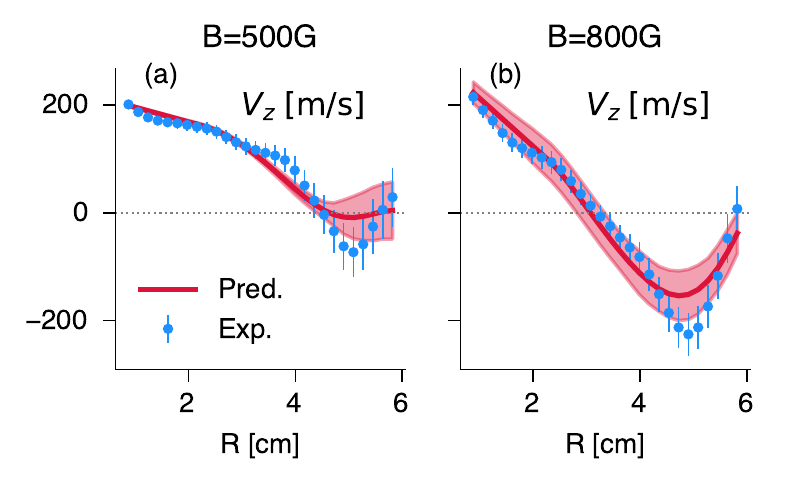}
    \caption{\label{fig:vzpred} Radial profiles of mean axial velocity predicted by force balance analysis (solid lines) and measured Mach probe (circles) at 500G (a) and 800G (b). Shaded area indicates the uncertainties of predicted $ V_{z} $ profile.}
\end{figure}

In order to determine if the observed changes in turbulent stress are responsible for the observed increase in sheared axial flow, an axial force balance analysis has also been performed.
The axial ion momentum equation is written as
\begin{equation}
\dfrac{1}{r}\dfrac{\partial }{\partial r} \left(r \langle \tilde{v}_{z}\tilde{v}_{r} \rangle\right) = -\dfrac{1}{m_i \langle n \rangle}\dfrac{\partial P_e}{\partial z} - \nu_{in} V_{z} +\dfrac{1}{r}\dfrac{\partial }{\partial r} \left( \mu_{ii} r \dfrac{\partial V_{z}}{\partial r} \right),
\label{eq:momentum}
\end{equation}
where the ion viscosity $ \mu_{ii} = \frac{6}{5} \rho_{i}^{2} \nu_{ii} \sim 5-10\,\rm m^{2}/s $ and ion-neutral collision frequency $ \nu_{in} = n_{\rm gas} v_{ti} \sigma_{in} \sim 3-6 \times 10^{3} \,\rm s^{-1}$ are estimated using density and ion temperature profiles \cite{Holland2006PRL195002,Thakur2016PoP82112}.
$ \mu_{ii} $ and $ \nu_{in} $ are likely to have small spatial variations, i.e., $ \mu_{ii} \propto n T^{-1/2}_{i} $ and $ \nu_{in} \propto T^{-1/2}_{i} $. 
Here, we assume the neutral pressure is radially uniform and the neutral temperature is approximated by the ion temperature which has been measured using LIF techniques \cite{Thakur2016PoP82112}.
This assumption gives the smallest estimate of neutral gas density depletion in the core and thus higher ion-neutral frictional dissipation.
But in this experiment, the ion-neutral drag dissipation is much smaller than the ion-ion collisional dissipation, and the force balance analysis is not sensitive to neutral profiles.
A no-slip boundary condition is also imposed due to the observations in Fig. 1(d), i.e., $ V_z \rightarrow 0 $ at $ r = 6 $ cm.
Taking the measured profiles of the Reynolds stress and the axial pressure gradient shown in \cref{fig:profiles}, we can solve \cref{eq:momentum} for $ V_z $ using a finite difference method.
As shown in \cref{fig:vzpred}, the calculated results (red curves) are in agreement with the mean axial ion flow profiles measured by the Mach probe (blue circles).
This result shows that the weak shear flow found at 500G is consistent with the weak axial equilibrium pressure gradient, while the stronger shear flow at 800G is consistent with the observed turbulent stress
Similar comparisons have been carried out throughout the dataset ($ B = 500 - 1000 $ G), and the agreement between measured and calculated axial flow profiles is found across a range of $ B $ fields.

\begin{figure}
    \centering
    \includegraphics[width=2.8in]{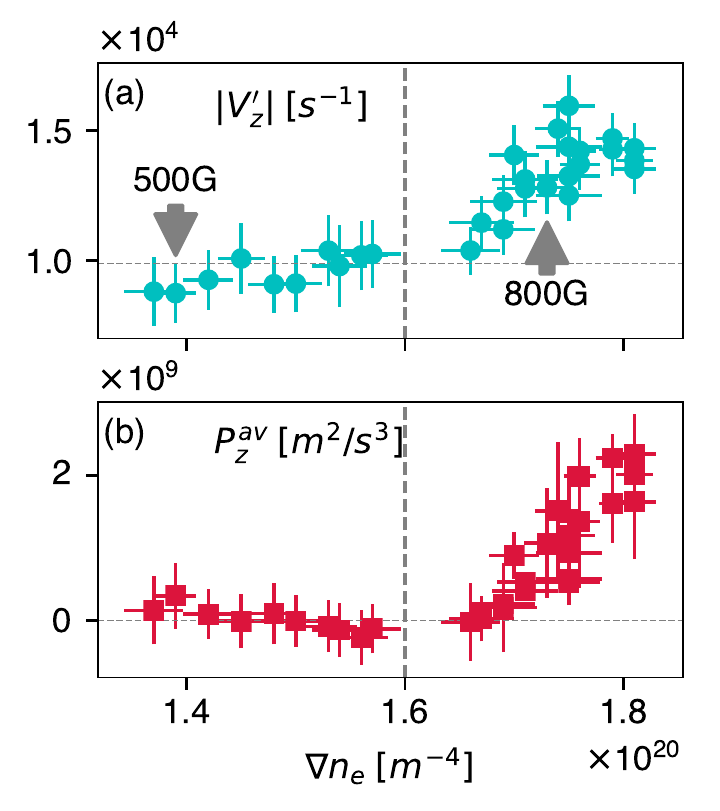}
    \caption{\label{fig:gradnescaling} The magnitude of axial flow shearing rate  $\left|\partial_r V_{z}\right|$ (a) and the volume-averaged axial Reynolds power $\mathcal{P}_{z}^{av}$ (b) plotted against the density gradient $ \nabla n_e $. Datasets are obtained by shot-by-shot $ B $ field scan.}
\end{figure}

Using a steady-state shot-by-shot $ B $ field scan, we illustrate the link between $\nabla n$, the turbulent flow drive, and the macroscopic intrinsic flow. The Reynolds power, $ \mathcal{P}_{z}^{Re} = - \langle V_z \rangle \partial_r \langle\tilde{v}_{z}\tilde{v}_{r}\rangle $, gives the rate of work performed by the turbulent fluctuations on the mean axial flow \cite{Tynan2016PPaCF44003} at any point in the plasma.  The volume-averaged Reynolds power, $ \mathcal{P}_{z}^{av} = \int - \langle V_z \rangle \partial_r \langle\tilde{v}_{z}\tilde{v}_{r}\rangle \,rdr / \int rdr $ where $ 1 < r < 5 $ cm to then gives the overall strength of the turbulent flow drive.
Below a threshold value of $B\approx 650$ G, the turbulent flow drive is small and the seed axial flow shear is driven by the axial pressure drop as shown earlier in \cref{fig:vzpred}(a), and varies at best weakly with $\nabla n$. 
The magnitude of axial flow shearing rate, $|V_{z}^{\prime}| = |\partial_{r}V_{z}|$, then increases sharply when the density gradient exceeds a critical value, $ \nabla n > 1.6 \times 10^{20} \,{\rm m^{-4}} $ (\cref{fig:gradnescaling}(a)) corresponding to $B\approx 650$ G.  
Further increases in $\nabla n$ associated with increased B then are associated with strong increased in flow shear and Reynolds power.  
These observations show that both the axial shear flow and its turbulent drive increase as $ \nabla n $ increases.

The critical density gradient behavior shown in \cref{fig:gradnescaling} has been reported previously \cite{Yan2010PoP12302}, which is in agreement with numerical simulations of the coupled drift wave turbulence--zonal flow system.
Another possible mechanism regarding this transition is that the small plasma radius in CSDX sets a lower bound for possible $ k_\perp \rho_s $ values, and thus prevents the growth of $ m=1 $ fluctuations at lower $ B $ fields.
Raising the $ B $ field lessens this geometry constraint and allows the onset of stronger turbulence.

The above observations suggest that an increase in $ \nabla n $ should drive larger residual stress and thus stronger intrinsic force, which then results in larger sheared flow.  Here, we derive a relation between the residual stress and the density gradient, and account for the symmetry breaking in this relation.
The fluctuating parallel ion flow evolves according to
\[
\frac{d\tilde{v}_{z}}{d t}=-c_{s}^{2}\nabla_{z}\left[\frac{e\tilde{\phi}}{T}+\frac{\tilde{P}}{P_{0}}\right]-\tilde{v}_{r}\frac{\partial V_{z}}{\partial r},
\]
where $c_{s}$ denotes the sound speed, $\tilde{v}_r$ is the eddy radial velocity, $\tilde{P}$ is the pressure fluctuation, and $\tilde{\phi}$ is the potential fluctuation.
Also, $V_z \nabla_z \tilde{v}_z$ is neglected due to small parallel wavenumber of the fluctuations $k_z \ll \omega/V_z$.
For drift wave turbulence with adiabatic electrons, one has $e\tilde{\phi}/T\sim\tilde{n}/n_{0}$, and $\tilde{P}/P_{0}\sim\tilde{n}/n_{0}$ as temperature fluctuations are negligible in this experiment \cite{Burin2005PoP52320,Thakur2014PSSaT44006}. 
The axial flow fluctuation can then be written as
$
\tilde{v}_{z}\approx-\sigma_{vT}\frac{c_{s}^{2}\tau_{c}}{L_{z}}\frac{\tilde{n}}{n_{0}} - \tilde{v}_{r}\tau_{c}\frac{\partial V_{z}}{\partial r}.
$
Here, $\tau_{c}$ is the eddy correlation time, $L_{z}$ is the characteristic axial dimension, and $\sigma_{vT}$ is a coefficient for acoustic coupling.
Using a mixing length model for the density fluctuation, $\tilde{n}/n_{0} \sim \frac{l_{c}}{n_{0}}\left|\frac{\partial n_{0}}{\partial r}\right|$, where $l_c \sim \tilde{v}_r \tau_c$ denotes the mixing length, one obtains
$$
\tilde{v}_{z}\approx-\sigma_{vT}\frac{c_{s}^{2}l_{c}^{2}}{L_{z}\tilde{v}_{r}}\frac{1}{n_{0}}\left|\frac{\partial n_{0}}{\partial r} \right| - \tilde{v}_{r}\tau_{c}\frac{\partial V_{z}}{\partial r}.
$$
After multiplying by $ \tilde{v}_{r} $ and taking an ensemble average, one obtains the expression for the total turbulent stress.
This consists of two parts, a turbulent diffusive flux proportional to the velocity shear and a residual term driven by the density gradient,
\[
\left\langle \tilde{v}_{r}\tilde{v}_{z}\right\rangle =-\chi_{z}\frac{\partial V_{z}}{\partial r}-\sigma_{vT}\frac{c_{s}^{2}\left\langle l_{c}^{2}\right\rangle }{L_{z}}\frac{1}{n_{0}}\left|\frac{\partial n_{0}}{\partial r}\right|.
\]
Here $\chi_{z}\sim \left\langle \tilde{v}_{r}^{2}\right\rangle \tau_{c}$ is the turbulent viscosity. 
The coefficient $\sigma_{vT}$ accounts for the efficiency of the density gradient in driving the residual stress, $\Pi_{rz}^{\textrm{Res}}$, via symmetry breaking.
In particular, $ \sigma_{vT} $ accounts for the spectral correlation $ \left\langle k_{z}k_{\theta}\right\rangle =\sum_{{\bf k}}k_{z}k_{\theta}\left|\hat{\phi}_{{\bf k}}\right|^{2}/\sum_{{\bf k}}\left|\hat{\phi}_{{\bf k}}\right|^{2} $, which encodes the broken symmetry of the turbulence.  Because all other terms can be measured in our experiment,
$\sigma_{vT}$ can be obtained by a least-square fit.

\begin{figure}
    \centering
    \includegraphics[width=2.5in]{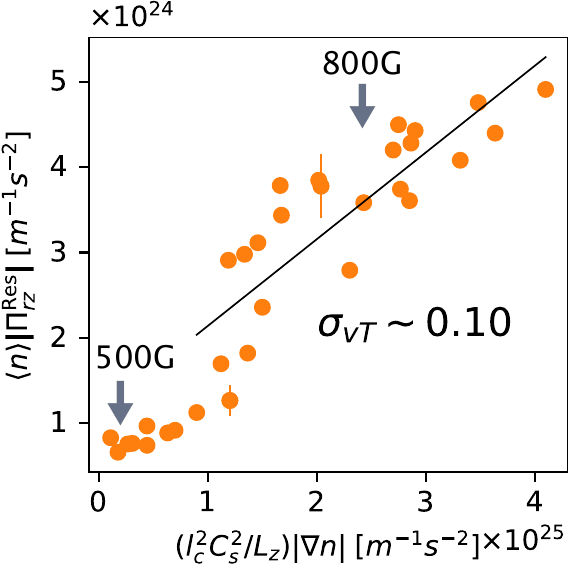}
    \caption{\label{fig:vT} Comparison between magnitudes of residual stress and normalized density gradient. The coefficient, $\sigma_{vT}$, is estimated to be about 0.10 by a least-square fit.}
\end{figure}

The residual stress $\Pi_{rz}^{\textrm{Res}}$ was synthesized from the measured Reynolds stress and the diffusive stress inferred from experimental measurements; the result was shown earlier in \cref{fig:profiles}(c).
As shown in \cref{fig:vT}, at smaller density gradient, the magnitude of residual stress, $\left|\Pi_{rz}^{\textrm{Res}}\right|$, is small and almost independent of the normalized density gradient.
At larger $ \nabla n $, $\left|\Pi_{rz}^{\textrm{Res}}\right|$ increases in proportion to the normalized density gradient, with a slope $\sigma_{vT}\approx 0.10$.
Here, $ \left|\Pi_{rz}^{\textrm{Res}}\right| $ is volume-averaged in the range of $ 1<r<5 $ cm.
This finding strongly supports the hypothesis that the residual stress is driven by the density gradient when the gradient exceeds a critical value.  The emergence of a finite $\sigma_{vT} \approx 0.1$ then indicates a symmetry breaking mechanism that emerges at higher $\nabla n$.

\begin{figure}
    \centering
    \includegraphics[width=3.5in]{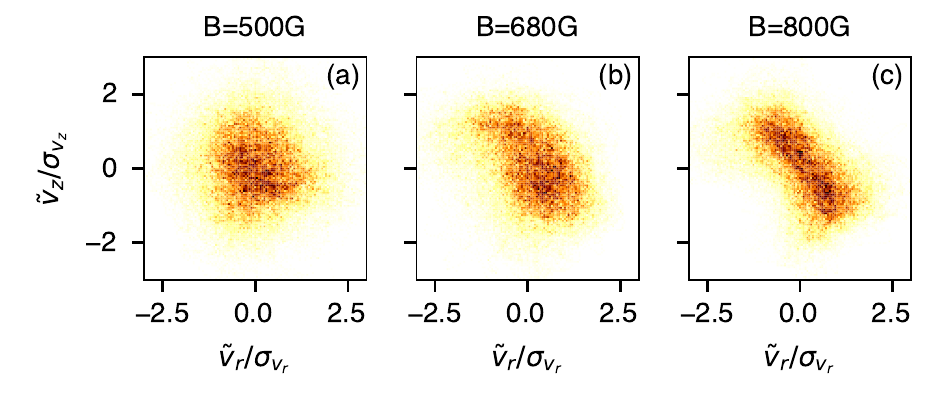}
    \caption{\label{fig:jointpdfs} Joint PDF of radial and axial velocity fluctuations, $\mathsf{P}\left(\tilde{v}_{r},\tilde{v}_{z}\right)$, at different magnetic fields at $r \approx 3$ cm. Normalization is the standard deviations.}
\end{figure}

The development of residual stress requires symmetry breaking in \textbf{k}-space \cite{Diamond2013NF104019}, i.e., $ \left< k_{z} k_{\theta} \right> = \sum_{\textbf{k}} k_{z} k_{\theta} \left|\hat{\phi}_{\textbf{k}}\right|^{2} / \sum_{\textbf{k}} \left|\hat{\phi}_{\textbf{k}}\right|^{2} \neq 0 $.
The symmetry breaking can be assessed by investigating the joint probability density function (PDF) of radial and axial velocity fluctuations, $ \mathsf{P}\left(\tilde{v}_{r},\tilde{v}_{z}\right) $.
In CSDX we have $ \tilde{v}_{z} \sim \nabla_{\parallel} \tilde{P} \sim k_z \tilde{\phi} $ and $ \tilde{v}_{r} \sim k_{\theta} \tilde{\phi} $, due to the nearly adiabatic electron response and negligible temperature fluctuations.
By normalizing the velocity fluctuations using their standard deviations, $ \mathsf{P}\left(\tilde{v}_{r},\tilde{v}_{z}\right) $ represents the correlator $ \left< k_{z} k_{\theta} \right> $.
As shown in \cref{fig:jointpdfs}, the anisotropy of $\mathsf{P}\left(\tilde{v}_{r},\tilde{v}_{z}\right)$ grows with increasing $B$ field strength and $ \nabla n $.
The highly elongated $\mathsf{P}\left(\tilde{v}_{r},\tilde{v}_{z}\right)$ at higher $ \nabla n $ indicates increased asymmetry in $ \left< k_{z} k_{\theta} \right> $.
Since larger residual stress occurs at higher $ \nabla n $, we can therefore infer that this symmetry breaking is related to the emergence of finite residual stress.

Conventional models of the origin of symmetry breaking rely on effects of magnetic geometry \cite{Diamond2013NF104019,Guercan2007PoP42306,Garbet2010PoP72505,Peeters2011NF94027,Angioni2012NF114003},
and therefore are not applicable to zero magnetic shear cases, such as CSDX.
To address this question, a dynamical symmetry breaking mechanism has been proposed \cite{Li2016PoP52311}.
This mechanism does not require magnetic shear, and may also be relevant to intrinsic rotation in tokamaks with flat-$q$ or weak shear.
This model is derived from a collisional electron drift-wave system with axial momentum evolution.
The mean axial flow shear then introduces a frequency shift proportional to $ k_z k_{\theta} V_{z}^{\prime} $ in drift wave growth rate.
In our experiments, the seed axial flow shear is negative, $ V_{z}^{\prime} < 0 $, because $ V_{z}(r) $ is initially driven by the axial pressure drop and decreases from the core to the edge.
As a result, modes with $ \left< k_{z} k_{\theta} \right> < 0 $ grow faster than modes with $ \left< k_{z} k_{\theta} \right> > 0 $, leading to spectral imbalance, with predominance of the spectral intensity in quadrants II and IV of the $ k_{\theta}-k_{z} $ plane.
The predicted spectral imbalance, $\langle k_\theta k_z \rangle <0 $, is consistent with the tilted contour of $\mathsf{P}\left(\tilde{v}_{r},\tilde{v}_{z}\right)$ (\cref{fig:jointpdfs}(c)). 
As demonstrated in Ref.~\cite{Li2016PoP52311}, the spectral asymmetry results in a residual stress of the form $- \chi_z^\text{Res} V_z^\prime$, with $\chi_z^\text{Res}<0$, i.e., a negative-definite contribution to the total viscosity (i.e., 
$\langle \tilde{v}_r \tilde{v}_z \rangle = - \chi_z V_z^\prime + \Pi_{rz}^\text{Res}  = (- \chi_z + \left| \chi_z^\text{Res} \right|) V_z^\prime$). 
Then, $\left| \chi_z^\text{Res} \right|=\chi_z$ defines the threshold $\nabla n_0/n_0$ for onset of axial flow generation. 
Using Eq.~(36) of Ref.~\cite{Li2016PoP52311} for $\left| \chi_z^\text{Res} \right|$, it gives  $\nabla n_\textrm{crit} = \left(n_0 \alpha \omega_{*e}^2/\langle k_\theta k_z \rangle \rho_s c_s \right) \times \left( L_z / c_s^2 \tau_c \right) \sim 1.5 \times 10^{20}\,\mathrm{m^{-4}}$ in agreement with experiment.
Here, $\alpha = k_z^2 v_{te}^2 / \omega_{*e} \nu_{ei} \sim 1$ is the adiabaticity factor, the perpendicular turbulence scale length is $k_\theta \rho_s \sim 1.5$, the eddy correlation time is $\tau_c \sim 6 \times 10^{-5}\,\mathrm{s}$, and $\sigma_{vT} = \langle k_\theta k_z \rangle / \langle k_\theta^2\rangle \sim 0.1$. 

In summary, in this study detailed measurements of axial flows and turbulent Reynolds stresses have been performed in cylindrical plasmas without magnetic shear.
As the density profile steepens, Reynolds stress develops and in turn drives a sheared mean axial flow.
Both axial flow shearing rate and the turbulent Reynolds power increase with density gradient.
The magnitude of residual stress also scales with the density gradient.
$ \mathsf{P}\left(\tilde{v}_{r},\tilde{v}_{z}\right) $ becomes highly tilted and anisotropic at higher $\nabla n$, indicating an asymmetry in the spectral correlator $\left\langle k_{\theta}k_{z}\right\rangle$.
This symmetry breaking in \textbf{k}-space implies a finite residual stress observed at higher $ \nabla n $, and is consistent with a model of dynamical symmetry breaking in the turbulence.
These findings constitute the first demonstration of the causal link of spectral symmetry breaking in drift wave turbulence to the development of a non-diffusive, residual stress, and ultimately to the onset of intrinsic axial shear flow.

\section{Acknowledgments}
The authors thank Dr.~Z.B. Guo for helpful discussions.
We also thank one of the referee for pointing out another mechnism of turbulence development.
This work was supported by the Office of Science, U.S. Department of Energy under Contract Nos.~DE-FG02-07ER54912 and DE-FG02-04ER54738.

\bibliography{axial-flow-refs}
\end{document}